\documentclass[preprintnumbers]{revtex4}

\usepackage{graphicx}
\def\ie{{\it i.e.}}
\def\eg{{\it e.g.}}
\def\etc{{\it etc}}

\def\mpl{\ifmmode \overline M_{Pl}\else $\overline M_{Pl}$\fi}
\def\to{\rightarrow}

\begin{document}
\bibliographystyle{revtex}

\preprint{SLAC-PUB-9632}

\title{{Contact Interaction Searches at the Linear Collider: Energy, 
Luminosity and Positron Polarization Dependencies}\footnote{Summary of a talk 
given at the American Linear Collider Workshop, the University of Texas, 
Arlington, Jan. 9-11, 2003.}}

\author{Thomas G. Rizzo}
\email[]{rizzo@slac.stanford.edu}
\affiliation{Stanford Linear Accelerator Center, 
Stanford University, Stanford, California 94309 USA}

\date{\today}

\begin{abstract}
Many types of new physics can lead to contact interaction-like modifications 
in $e^+e^-$ processes below direct production threshold.  
This report summarizes a survey of contact interaction search reaches at the 
Linear Collider as functions of energy, luminosity and positron polarization. 
The various tradeoffs between these quantities in such searches are examined 
in detail.  
\end{abstract}

\maketitle

\section{Introduction}

It is generally expected that new physics beyond the Standard Model(SM) will 
manifest itself at future colliders that probe the TeV scale such as the LHC 
and the Linear Collider(LC). This new physics(NP) may appear either directly, 
as in the case of new particle production, \eg, SUSY or Kaluza-Klein 
resonances, or indirectly through 
deviations from the predictions of the SM. In the case of indirect discovery 
the effects may be subtle and many different NP scenarios may lead to the 
same or very similar experimental signatures. 

Perhaps the most well known example of this indirect scenario in a 
collider context would be the observation of deviations 
in, \eg, various $e^+e^-$ cross sections due to apparent contact 
interactions. There are many very different NP scenarios 
that predict new particle exchanges which can lead to contact interactions 
below direct production threshold; 
a partial list of known candidates is: compositeness{\cite {elp}}, 
a $Z'$ from an extended electroweak 
gauge model{\cite {e6,zp}}, scalar or vector  
leptoquarks{\cite {e6,lq}, $R$-parity violating sneutrino($\tilde \nu$) 
exchange{\cite {rp}}, scalar or vector 
bileptons{\cite {bl}}, graviton Kaluza-Klein(KK) 
towers{\cite {ed,dhr}} in extra dimensional models{\cite {add,rs}}, 
gauge boson KK towers{\cite {ed2,dhr}}, and even string 
excitations{\cite {se}}. Of course, there may be many other 
sources of contact interactions from NP models as yet undiscovered, as was the 
low-scale gravity scenario only a few years ago.

The purpose of this paper is to overview how contact interaction search 
reaches are influenced by changes in the LC center of mass energy, integrated 
luminosity and positron polarization{\cite {tdr}}. 
To be specific we will limit our  
discussion to the processes $e^+e^- \to \bar ff$ and   
to four of the scenarios listed above: new $Z'$'s, gauge KK towers in the 
5-dimensional version of the SM(5DSM), graviton 
exchange in the ADD model and compositeness. We will at first 
consider the following 
center of mass energies: $\sqrt s=0.5,0.8,1.0,1.2$ and 1.5 TeV and luminosities 
in the range $0.1 \leq L \leq 3~ab^{-1}$ and then generalize so that we may 
interpolate among these cases. Assuming an $e^-$ polarization of 
$80\%$ we initially consider only two possible polarizations for positrons: 
$P_+=0,60\%$ and later generalize to a continuum of values. 
In calculating errors, statistical uncertainties and those systematics 
arising from both 
polarization and luminosity uncertainties, $\delta P/P=0.003$ and $\delta L/L=
0.0025$ are employed. Initial state radiation but {\it no} beamstrahlung has 
been included and a symmetric low angle cut $\theta_{min}=100$ mrad 
has been imposed. 
Finite efficiencies for flavor tagging the final state leptons and quarks, 
$f=e, \mu, \tau, c,b,t$, are also included in the calculations. 
In performing fits we employ the following observables: 
the unpolarized total cross sections, $\sigma_f$, the 
unpolarized angular distributions, $1/\sigma_f ~d\sigma_f/d\cos \theta$, the 
left-right polarization asymmetries, $A_{LR}^f(\cos \theta)$ 
and the polarization of taus in 
the final state, $P_\tau$, including the effects of a finite efficiency. 
Comparisons between the predictions of the new physics 
models to those of the SM are determined by the $\chi^2$ of the fit which is 
controlled by a single parameter, a mass scale, in each case. The resulting $95\%$ 
CL 
bounds we obtain are consistent with those found in earlier analyses{\cite {tdr}}. 

However here it is not so much the bounds themselves that we are interested in but 
their {\it variation} as we change the values of $\sqrt s$, $L$ and $P_+$. 
For impatient readers the punchline of this analysis can be found in Section V and 
specifically in Figs. 8 and 9. Sections II-IV contain the justification for these 
later results and conclusions.

\section{New Gauge Bosons}

\begin{figure}[htbp]
\centerline{
\includegraphics[width=8.5cm,angle=90]{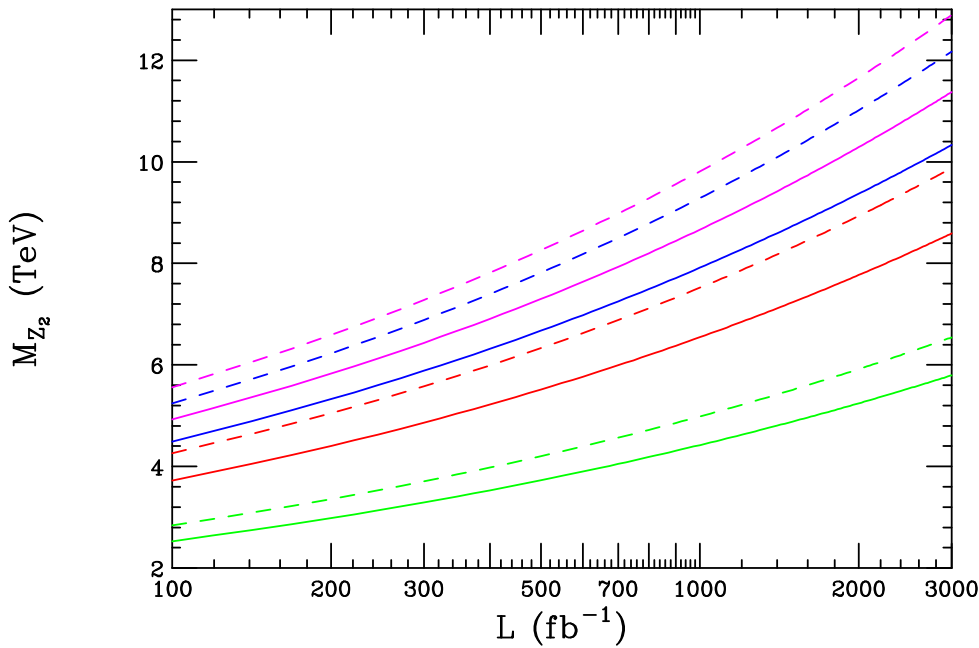}}
\vspace*{0.1cm}
\caption{$95\%$ CL search reach $M_{Z_2}$ 
for a $Z'$ at a $\sqrt s=500$ GeV LC when 
$P_+=0$(solid) or $60\%$(dashed) as a function of the integrated luminosity  
for $E_6$ models $\psi$(green), $\chi$(red), the LRM(blue) and the 
SSM(magenta).}
\label{fig1}
\end{figure}

Our first example of contact interactions is $Z'$ exchange below production 
threshold. There are a huge number of models of this kind{\cite {e6,zp}}.   
We will consider four specific but representative $Z'$ models: the models  
$\psi$ and $\chi$ based on $E_6$ GUTS{\cite {e6}}, the canonical Left-Right 
symmetric model(LRM) with equal left- and right-handed couplings, \ie, 
$\kappa=g_R/g_L=1${\cite {zp}} and 
the so-called Sequential Standard Model(SSM), which has 
a heavy copy of the SM $Z$ that is often 
used by experimenters to gauge $Z'$ sensitivity. The possibility of 
mixing between any of these new $Z'$ 
fields and the SM $Z$ will be neglected in the analysis and the deviations of 
the observables from their SM values for the final 
states $f=\mu, \tau, c,b$ and $t$ will be combined into a single overall fit. 
Fig.~\ref{fig1} shows a typical result from this analysis for the specific case 
of an LC with $\sqrt s=500$ GeV. 
We see immediately that the $Z'$ limits are quite model dependent as one might 
expect due to the wide variation in their couplings to the various SM fermions; 
note that reaches in the range $\sim 5-12\sqrt s$ seem rather generic.  
We also observe that the search reaches are in all cases 
relatively($\sim 10-20\%$) sensitive to the presence 
of positron polarization, but in a model-dependent manner. 
Although the various reach curves have slightly different 
slopes they are found to rise approximately{\cite {zp}} 
as $\sim L^{1/4}$ in all cases.
As we will discuss below this approximate scaling implies that the discovery 
reach is essentially statistics dominated. 

\begin{figure}[htbp]
\centerline{
\includegraphics[width=5.6cm,angle=90]{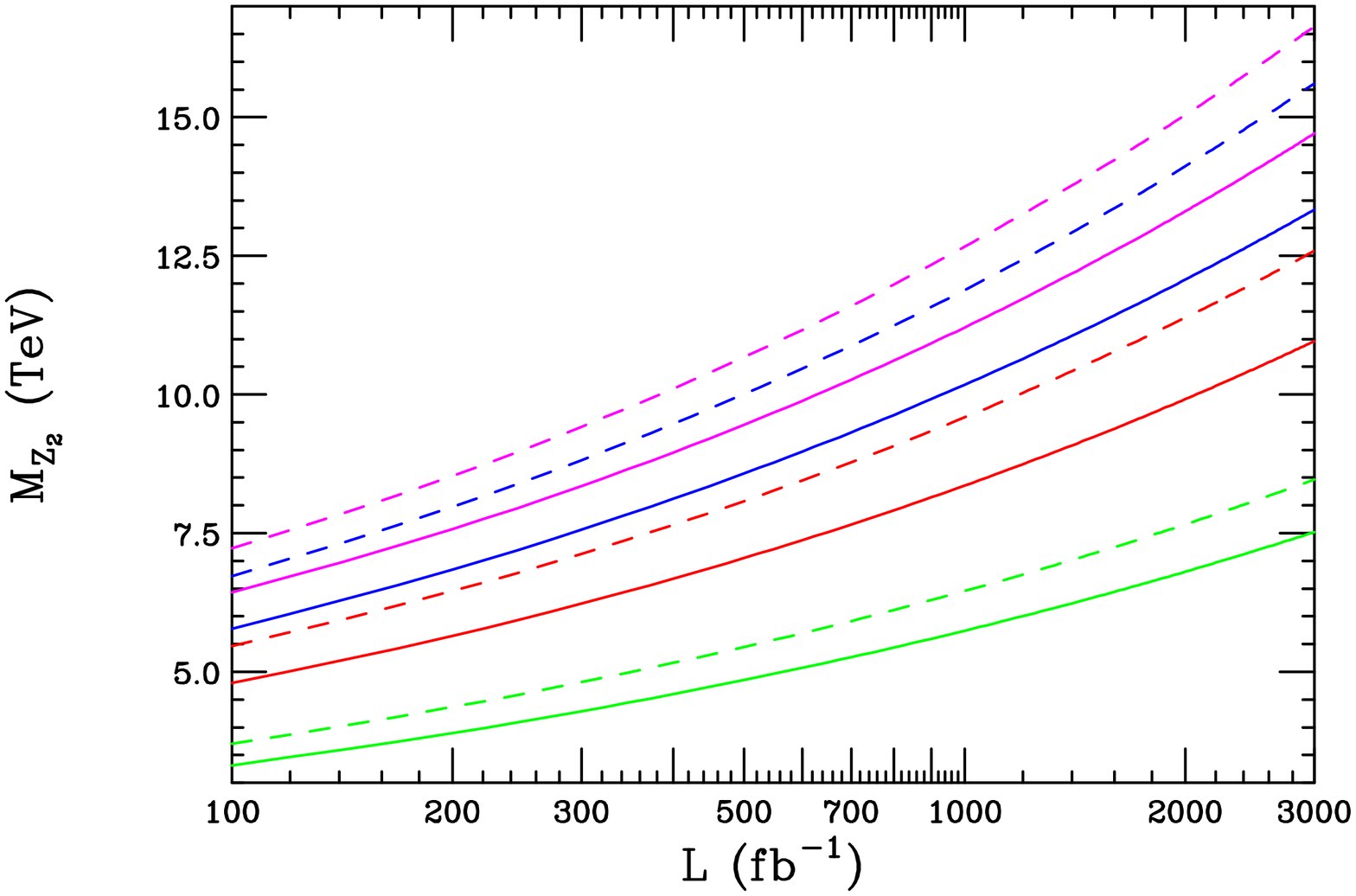}
\hspace*{5mm}
\includegraphics[width=5.6cm,angle=90]{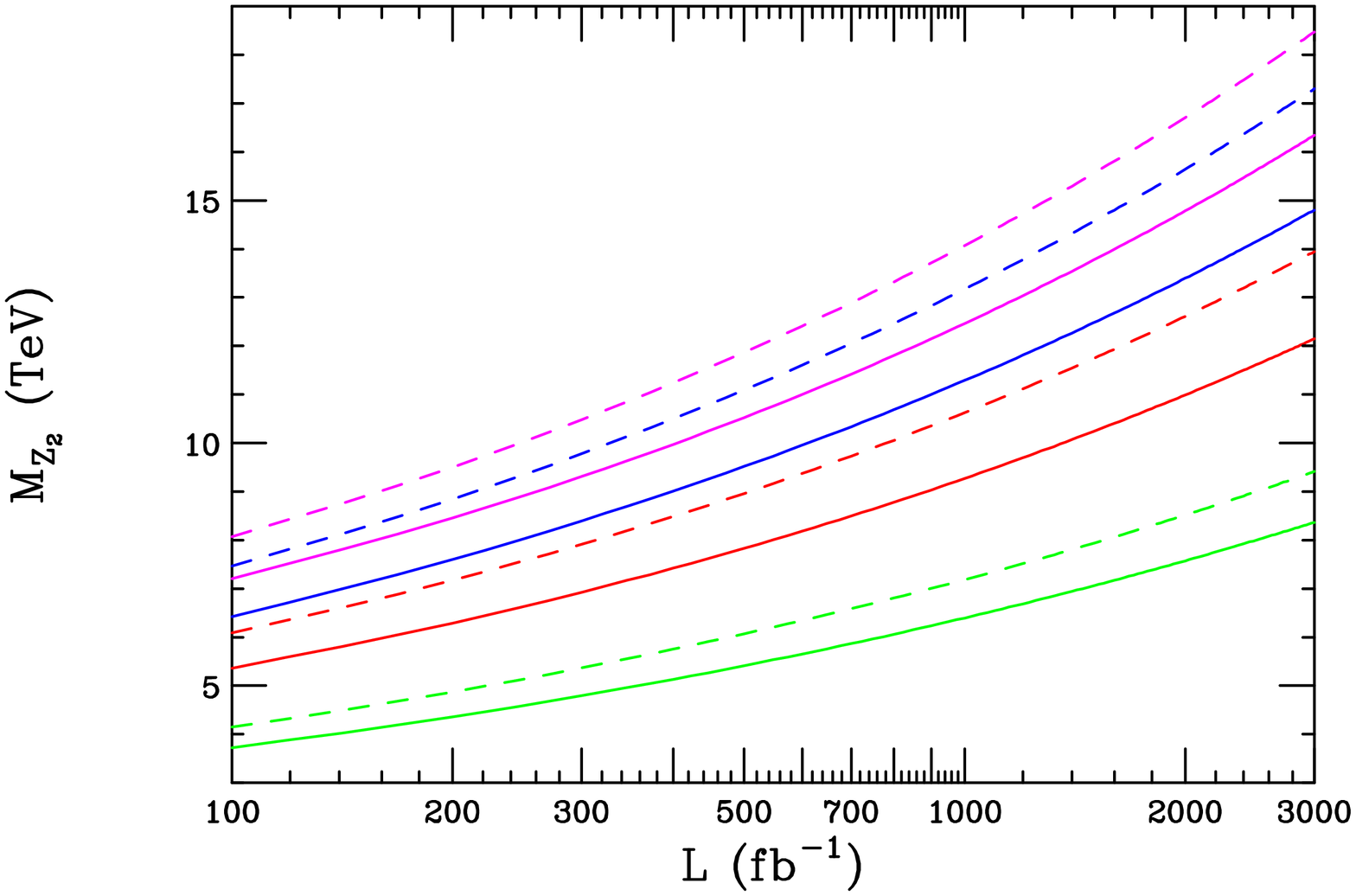}}
\vspace*{0.3cm}
\centerline{
\includegraphics[width=5.6cm,angle=90]{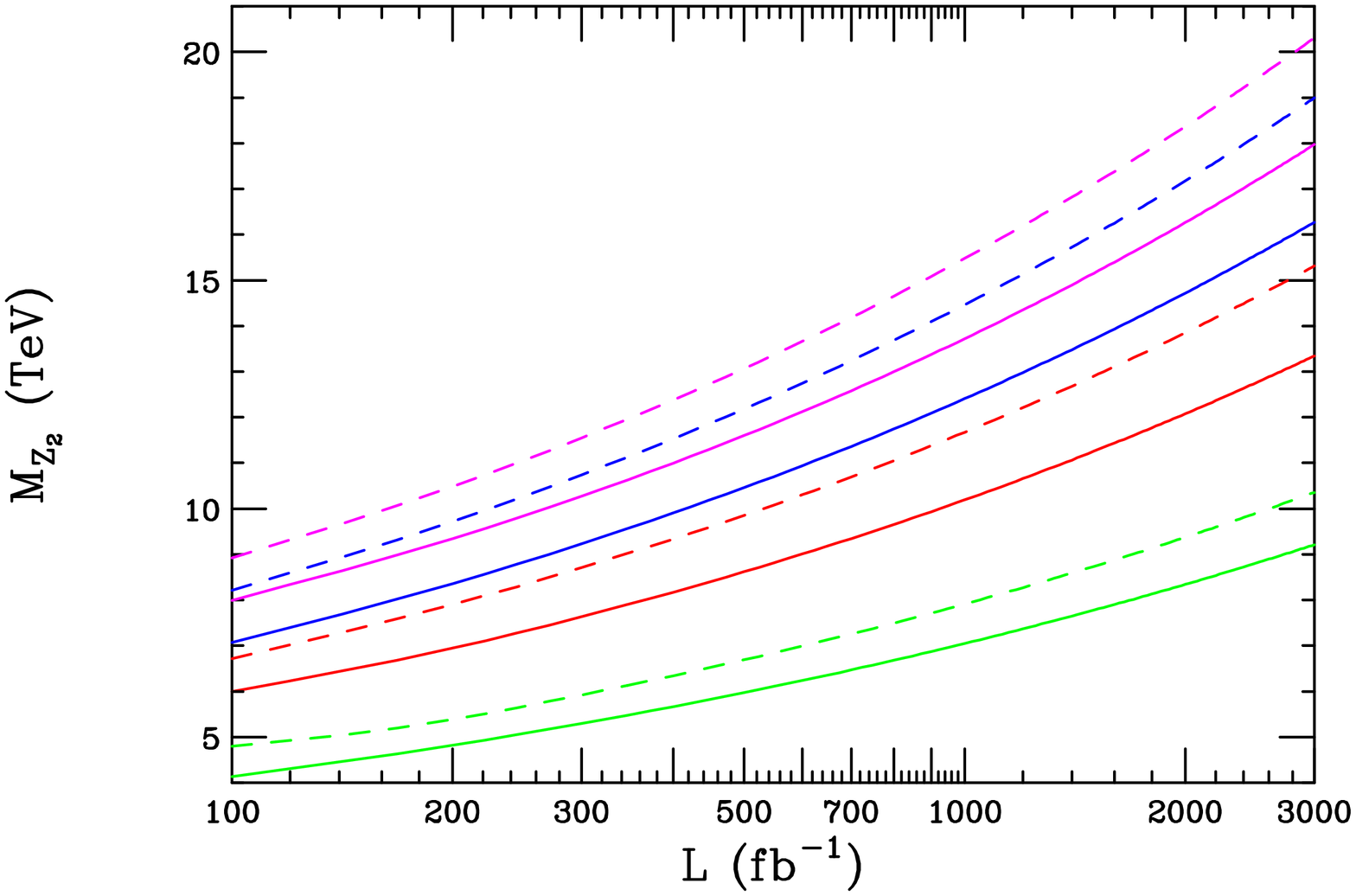}
\hspace*{5mm}
\includegraphics[width=5.6cm,angle=90]{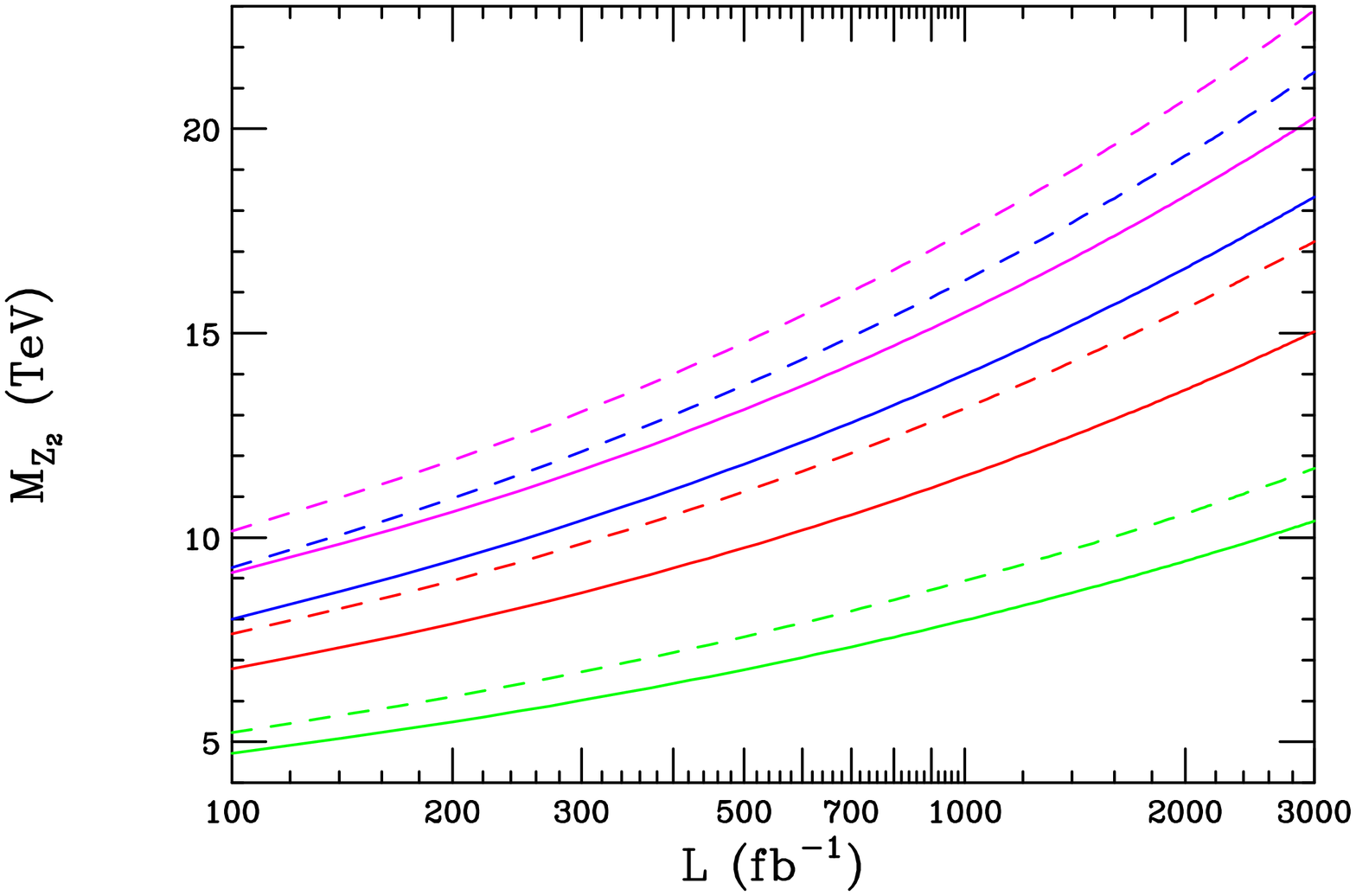}}
\vspace*{0.1cm}
\caption{Same as the previous figure but now for $\sqrt s=0.8$ TeV (upper left), 
1 TeV(upper right), 1.2 TeV(lower left) and 1.5 TeV(lower right).}
\label{fig2}
\end{figure}

What happens to these results as we vary $\sqrt s$? This is shown in detail in 
Fig.~\ref{fig2} for the same set of $Z'$ models. In order to compare the same 
model but at different values of 
$\sqrt s$, thus removing the coupling dependencies, we 
replot these results as shown in Fig.~\ref{fig3}. In both these figures we see 
that these are some small relative changes in the slopes as $\sqrt s$ is varied. 
Overall one sees that for a given model and 
fixed values of $L$ and $P_+$ the reach scales{\cite {zp}} approximately 
as $(\sqrt s)^{1/2}$, which again signals the dominance of statistical errors. 

\begin{figure}[htbp]
\centerline{
\includegraphics[width=5.6cm,angle=90]{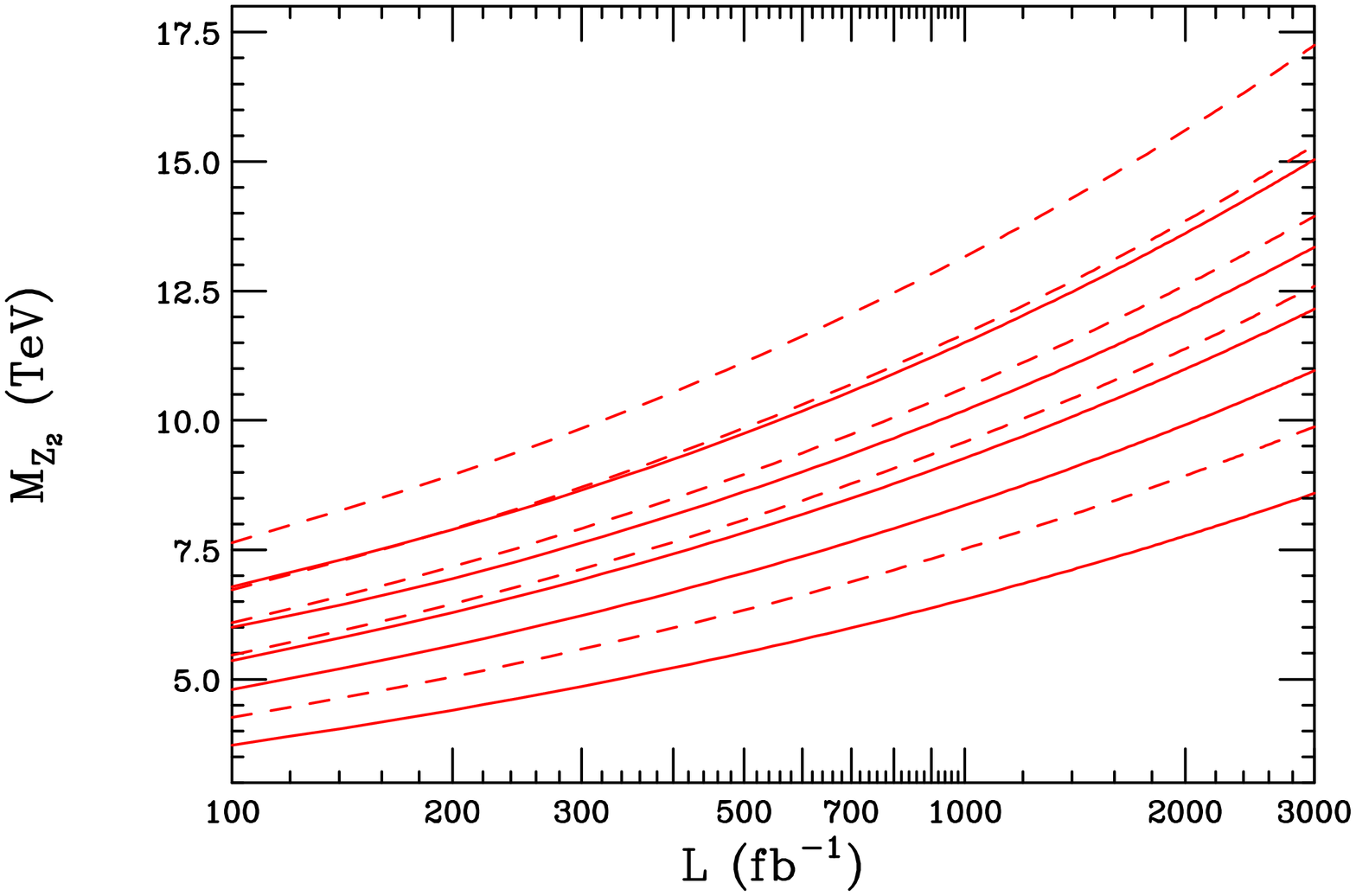}
\hspace*{5mm}
\includegraphics[width=5.6cm,angle=90]{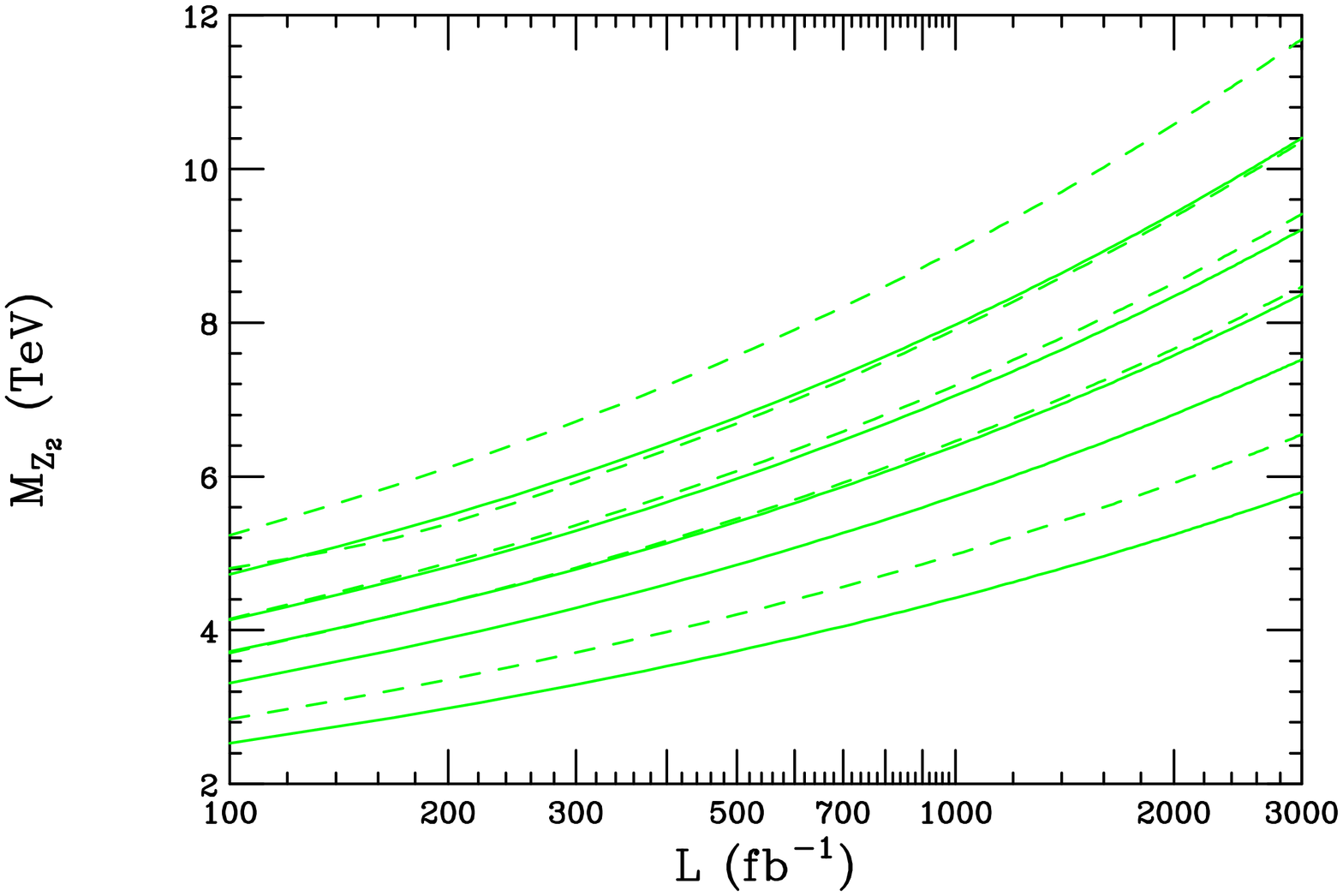}}
\vspace*{0.3cm}
\centerline{
\includegraphics[width=5.6cm,angle=90]{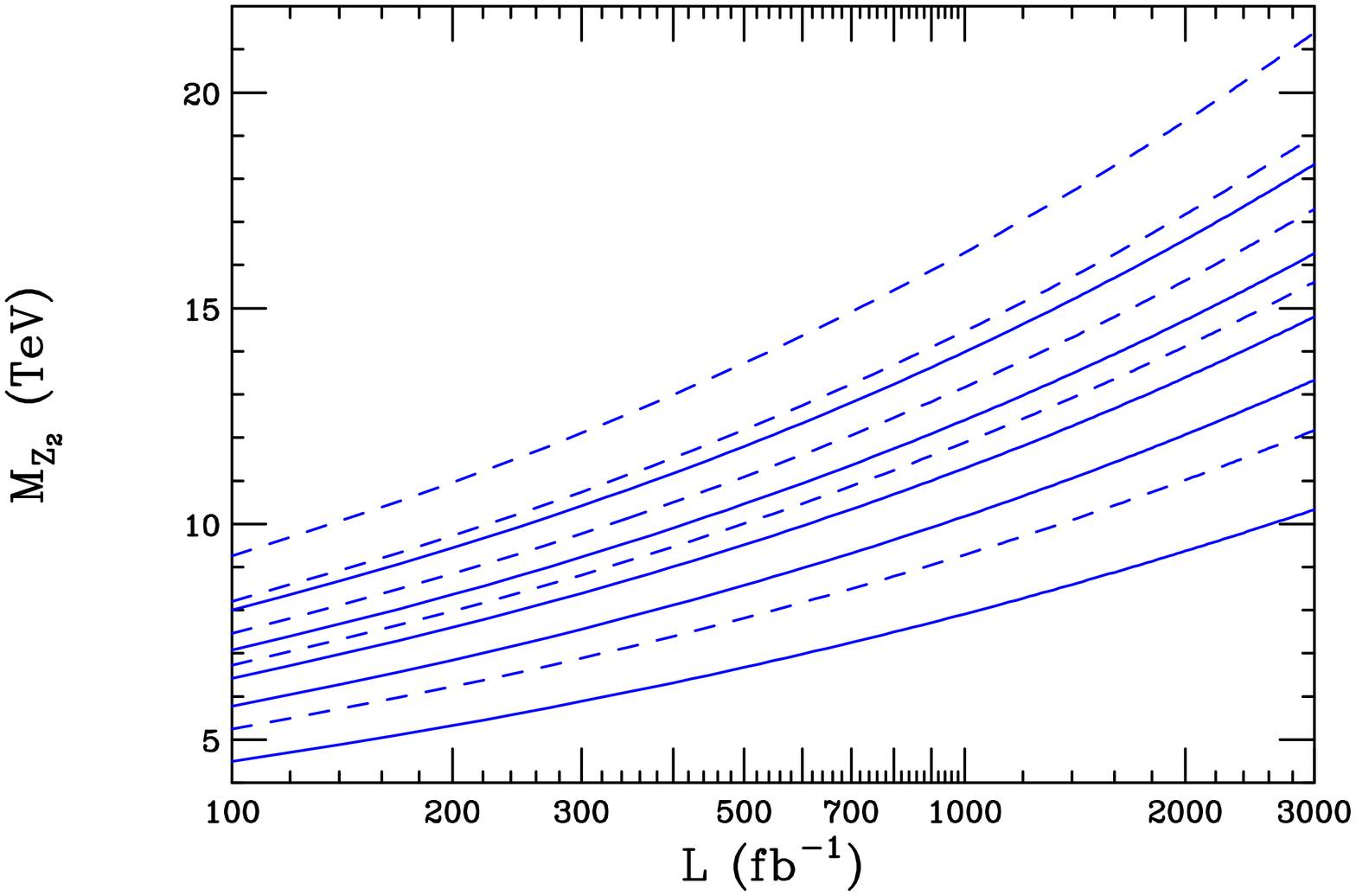}}
\vspace*{0.1cm}
\caption{$95\%$ CL search reaches for different $Z'$ models. In each plot 
there are a pair of curves (solid,dashed) for each center of mass energy 
corresponding to $P_+=(0,60)\%$. From bottom to top the curves correspond to 
$\sqrt s=0.5,0.8,1,1.2$ and $1.5$ TeV, respectively. 
The upper left(right) panel is for model 
$\chi(\psi)$ while the lower panel is for the LRM.}
\label{fig3}
\end{figure}

\section{Extra Dimensions}

We will investigate 
two models which display distinct signatures for extra dimensions. 
The first case we consider is the 5DSM in which the photon and $Z$(as well as the 
$W$ and gluon) of the SM have nearly equally spaced KK excitations with masses 
$M_n^2=n^2M_c^2+M_0^2$, where $M_0$ is the SM particle mass, $n=1,2,3,...$ and 
$M_c$ is the compactification scale. Based on analyses of precision measurements 
we expect $M_c$ to be in excess of 4-5 TeV{\cite {ed2}}. (We note however that 
there are variations of this model where KK gauge boson can exists which are 
significantly lighter than these bounds thus allowing for the possibility of 
their direct production at a TeV class LC.) The scale $M_c$ is thus 
essentially the mass of the first photon/$Z$ KK excitation. The reach for $M_c$ 
is expected to be significantly larger than, say, the reach for the mass of the 
SSM $Z'$, for several reasons. First, both the photon and $Z$ have KK excitations 
which perturb the various observables. Second, a whole tower of both states 
exists instead of a single state 
and, lastly, the couplings of all the states in both of the KK towers to the SM 
fermions are greater than those of the corresponding 
SM gauge fields by a factor of $\sqrt 2$. 
Fig.~\ref{fig4} shows the search reaches obtained in this case with values of 
$M_c$ as high as $\sim 20\sqrt s$ being probed. We see several things from this 
plot. First, the curves have very similar slopes though there is some 
variation as the values of $\sqrt s$ and $P_+$ are altered. However, to a very 
good 
approximation the reach in all cases scales as $\sim (sL)^{1/4}$ which signals 
that the bounds are statistics dominated. Second, increasing $P_+$ from 0 to $60\%$ 
leads to an increase in the reach for all $\sqrt s$ of approximately $\sim 12\%$. 
This is similar to what was seen in the case of new gauge bosons.

\begin{figure}[htbp]
\centerline{
\includegraphics[width=8.5cm,angle=90]{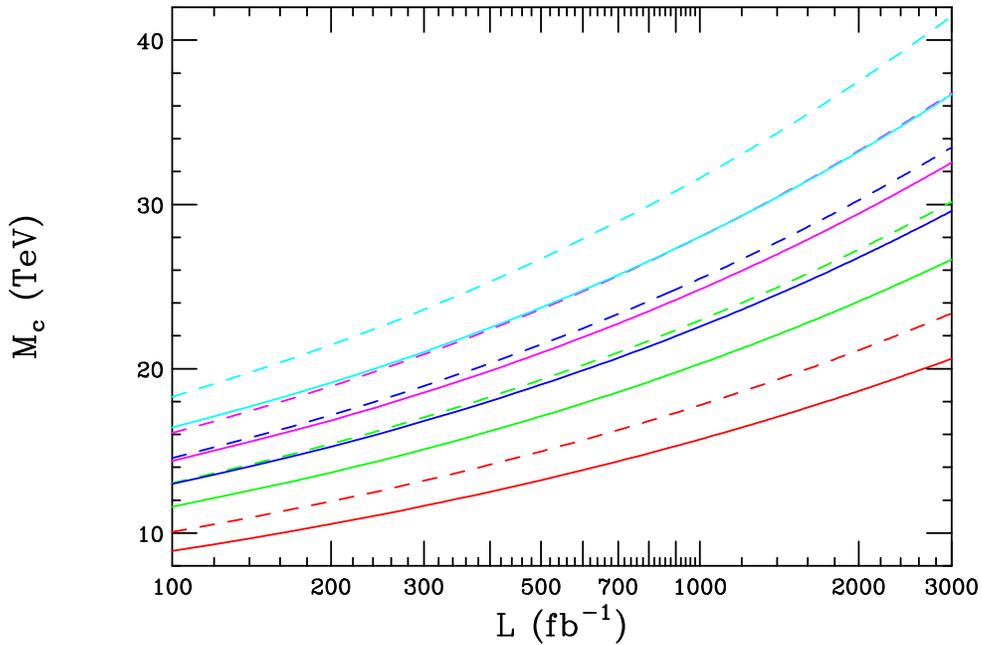}}
\vspace*{0.1cm}
\caption{$95\%$ CL search reach for the compactification scale of the 5DSM. 
There are a pair of curves (solid,dashed) for each center of mass energy 
corresponding to $P_+=(0,60)\%$. From bottom to top the curves correspond to 
$\sqrt s=0.5,0.8,1,1.2$ and $1.5$ TeV.}
\label{fig4}
\end{figure}

In the second extra-dimensional scenario, the ADD model, an almost continuum 
tower of KK gravitons is exchanged. The corresponding KK sum needs to be cutoff 
and is defined 
by a parameter $\Lambda_H$ if the scheme of Hewett is employed{\cite {ed2}}. 
Graviton exchange differs in an important way from the other contact interactions  
encountered above in that to lowest order it can be represented as an operator of  
dimension-8. $Z'$, 5DSM and compositeness effects all lead to dimension-6 
operators. We note that for dimension-$d$ operators the new effects due to the 
associated contact 
interactions scale as $\sim (\sqrt s/M)^{(d-4)}$, where $M$ is the relevant mass 
scale. Furthermore, a short analysis demonstrates that for dimension-$d$ operators 
the reach scales correspondingly as
\begin{equation}
\sim \Big[s^{(d-5)}L\Big]^{1/(2d-8)}\,,
\end{equation}
when statistical errors dominate. For the now familiar dimension-6 case this yields 
$\sim (sL)^{1/4}$ as discussed above but for the ADD model we obtain instead the 
result $\sim (s^3L)^{1/8}$. This $\sim L^{1/8}$ behavior for the search reach 
is an excellent approximation to what is observed in Fig.~\ref{fig5}; the growth in 
the reach for fixed $L$ is also consistent with the $\sim (\sqrt s)^{3/4}$ expectation 
assuming statistically dominated errors. The increase in the reach with $P_+$ 
going from 0 to $60\%$ is somewhat smaller than in the two previously examined 
cases here being only $\simeq 5\%$2. Note that as the dimension of the NP operator 
increases, changes in $L$ become far less important than changes in $\sqrt s$ 
in increasing search reaches. 

\begin{figure}[htbp]
\centerline{
\includegraphics[width=8.5cm,angle=90]{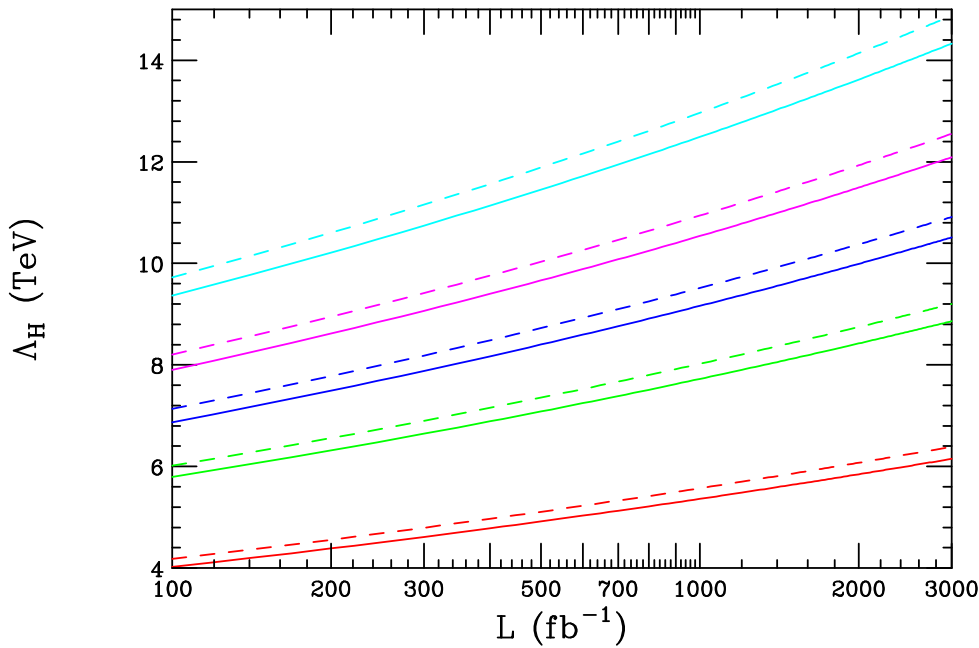}}
\vspace*{0.1cm}
\caption{Same as the previous figure but now for the cutoff scale in the ADD 
model employing the prescription of Hewett.}
\label{fig5}
\end{figure}

\section{Compositeness}

If the SM fermions are composite then they can exchange constituents during a 
scattering process which leads to new dimension-6 operators{\cite {elp}} and the 
corresponding contact interactions. Since different fermions may have different 
constituents and differing scales of compositeness, the simplest process to 
analyze in this case is Bhabha scattering since only electrons and positrons 
are involved. The contact interactions in this scenario can be classified according 
to the helicity structure of the two leptonic currents: LL, RR, LR, VV, AA, \etc ,
and the associated operator mass scale $\Lambda$. Here we will assume that these 
interactions constructively interfere with the SM $\gamma$ and $Z$ exchange 
contributions and that only one of these helicity structures dominates. 

\begin{figure}[htbp]
\centerline{
\includegraphics[width=7.6cm,angle=90]{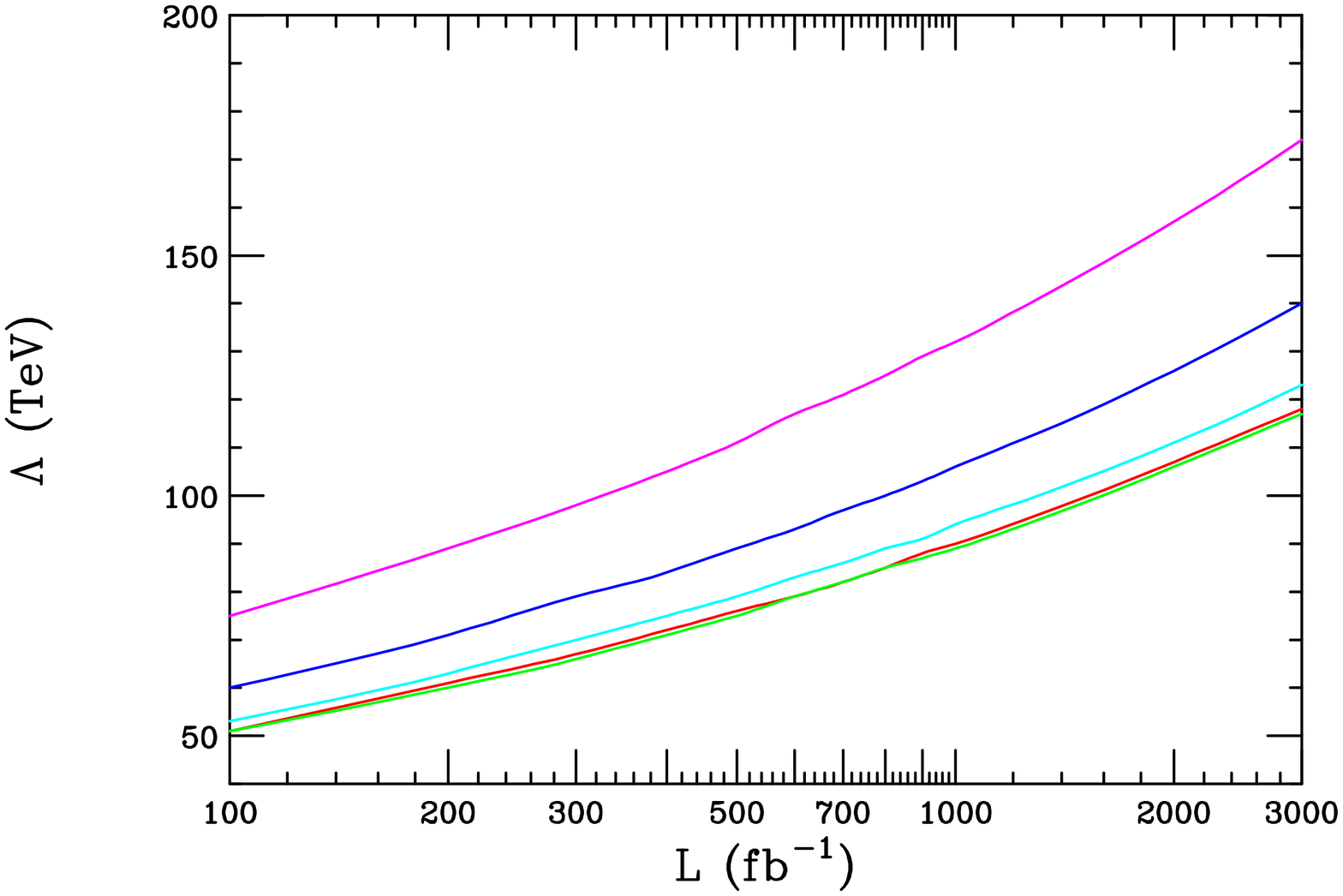}}
\vspace*{0.3cm} 
\centerline{
\includegraphics[width=7.6cm,angle=90]{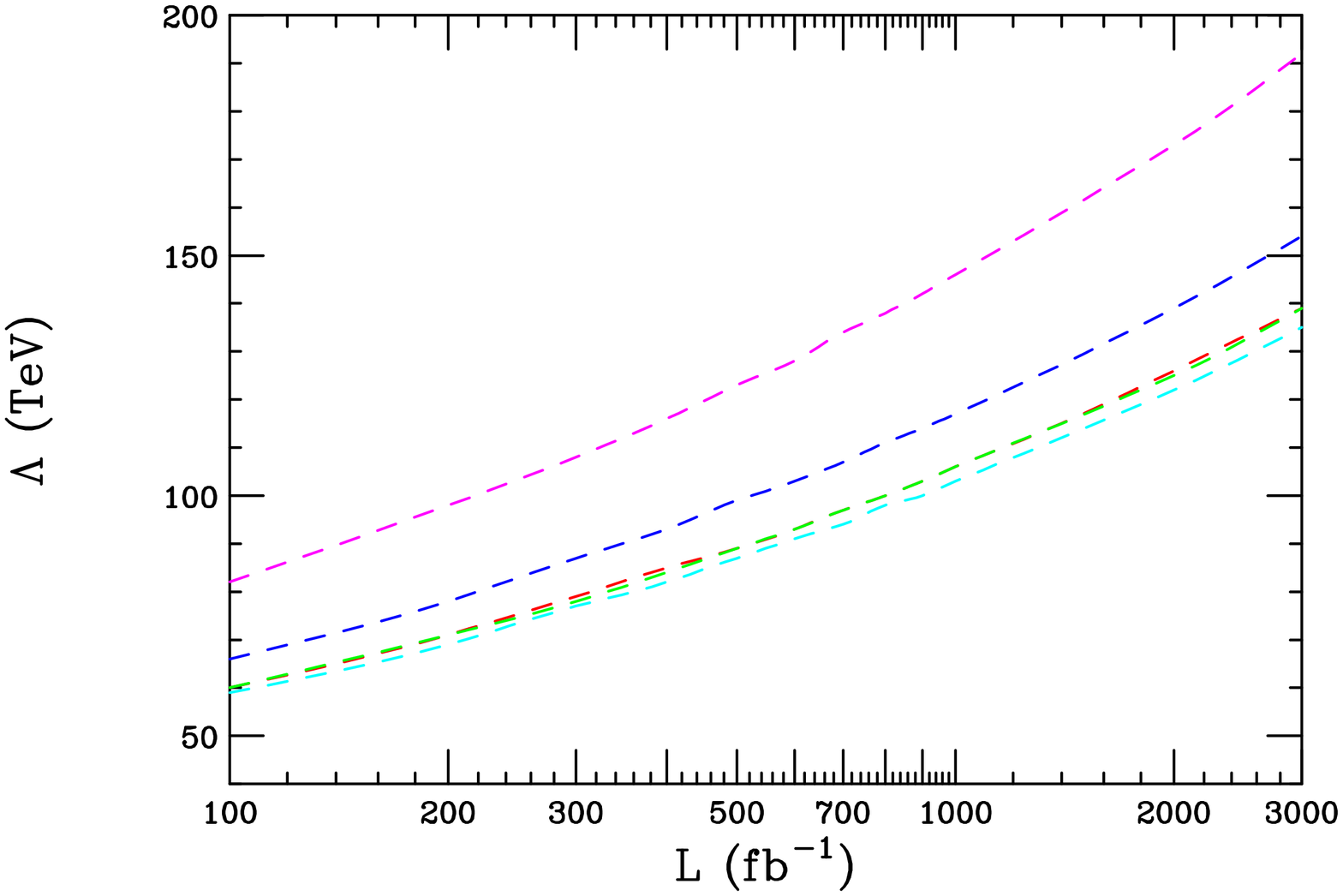}}
\vspace*{0.1cm}
\caption{Top(Bottom): $95\%$ CL lower bound on the compositeness scale at a 
$\sqrt s$=500 GeV 
LC arising from Bhabha scattering assuming $P_+=0(60)\%$. The 
red(green,blue,magenta,
cyan) curves are for LL(RR,LR,VV,AA) type interactions.}
\label{fig6}
\end{figure}

In the case of Bhabha scattering the $\theta_{min}$ cut can play a more 
important 
role than for the purely $s$-channel processes discussed above. The cut is 
essential in removing the purely photonic $t$-channel singularity in the 
forward direction. The overall 
event rate and hence to some extent the statistics is, however, sensitive 
to the particular value chosen for the cut. However, the very forward region, 
being so dominated by the pure photon exchange diagram, is not very sensitive 
to the existence of contact interaction contributions.  
Fig.~\ref{fig6} show the search reaches for these five 
helicity combinations at a $\sqrt s=500$ GeV LC both without and with positron 
polarization, respectively; note that the reach is quite 
sensitive to the helicity choice. The increased reach obtained with positron 
polarization is found to be helicity dependent but overall  
comparable to that obtained for the $Z'$ and 5DSM cases: $\sim 10-18\%$.  
Fig.~\ref{fig8} shows the corresponding 
$\sqrt s$ dependences of the search reaches. We find 
that again the reaches are statistically dominated so that they scale 
approximately as $\sim (sL)^{1/4}$ as they do for other dimension-6 
contact interactions.

\begin{figure}[htbp]
\centerline{
\includegraphics[width=5.6cm,angle=90]{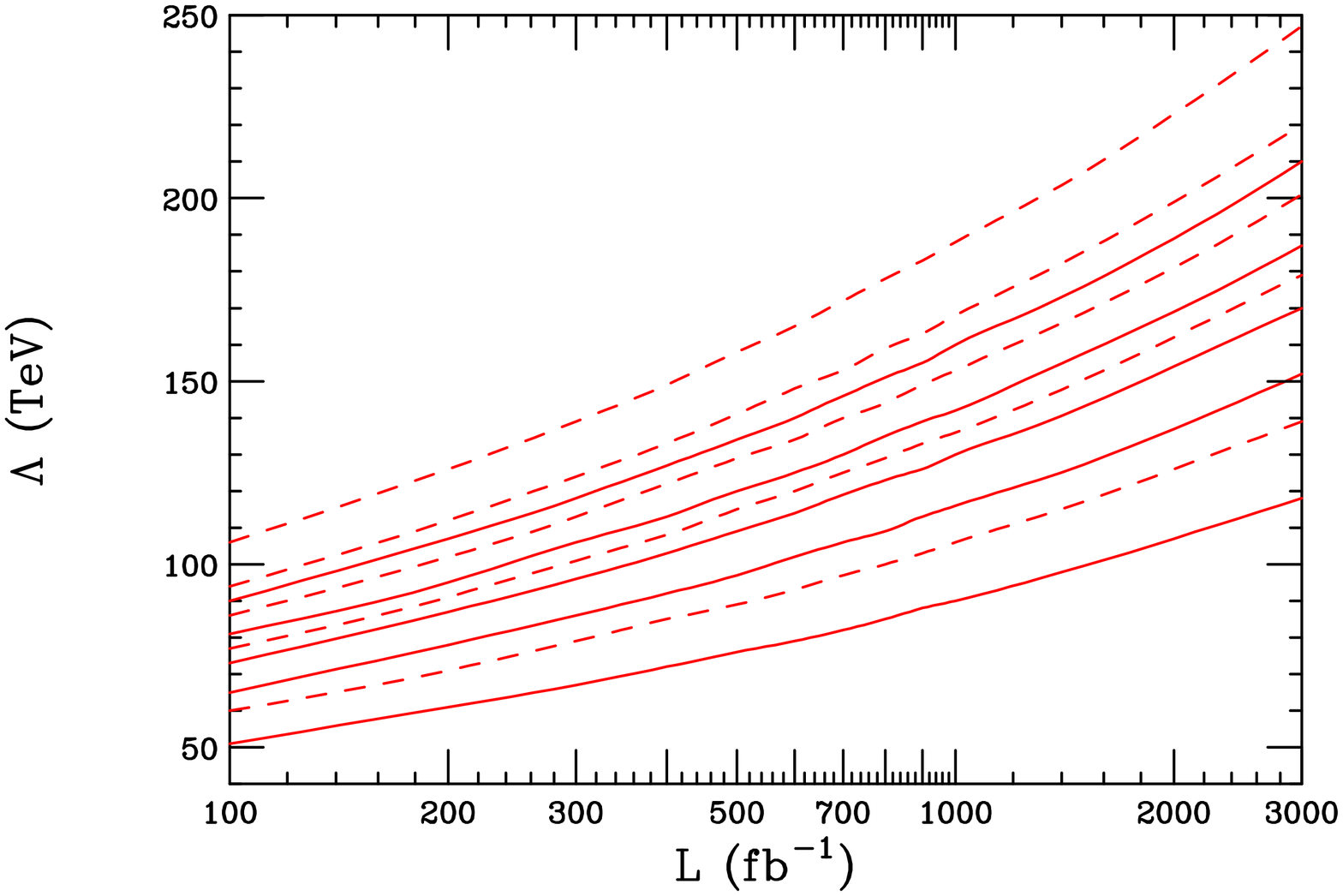}
\hspace*{5mm}
\includegraphics[width=5.6cm,angle=90]{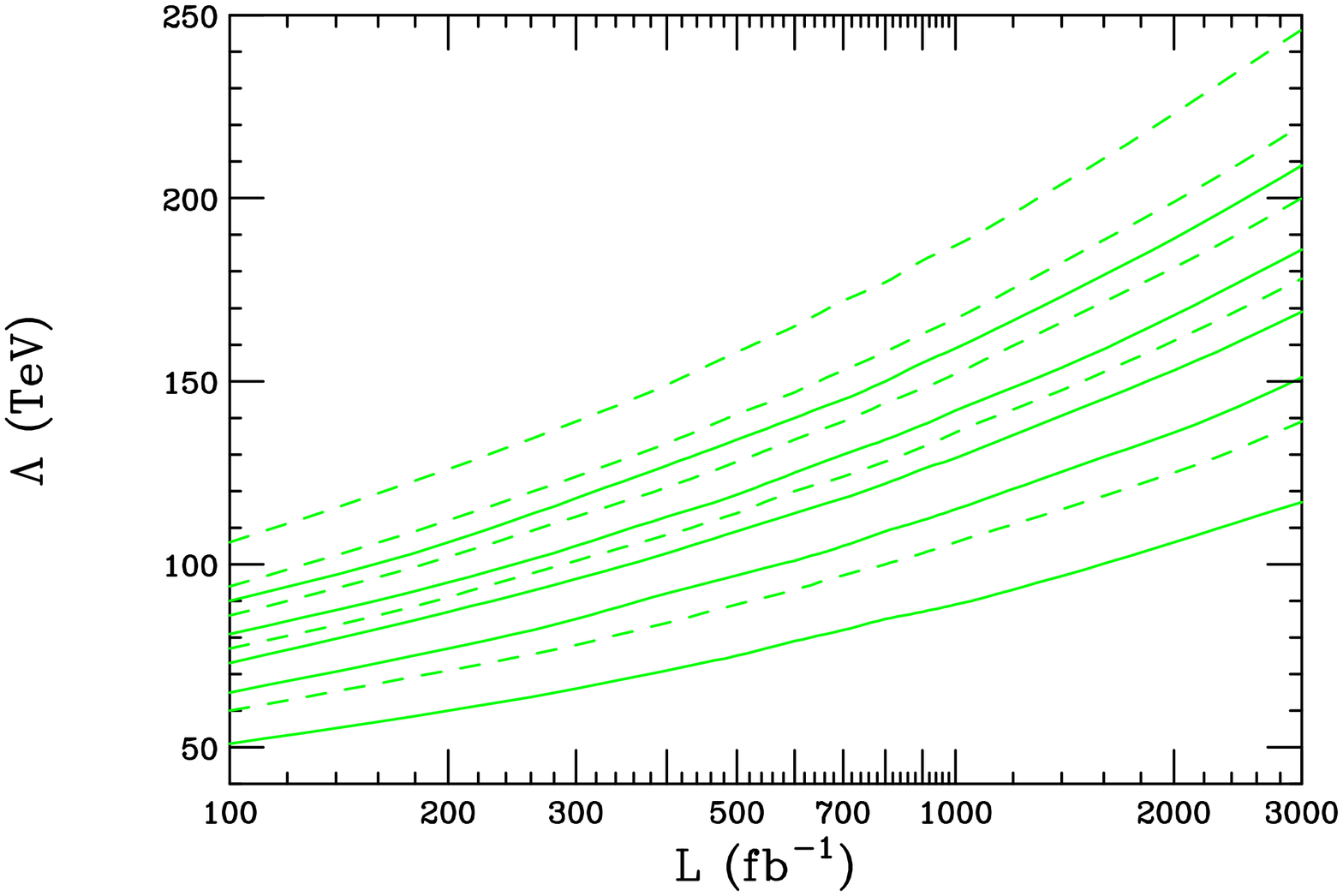}}
\vspace*{0.3cm}
\centerline{
\includegraphics[width=5.6cm,angle=90]{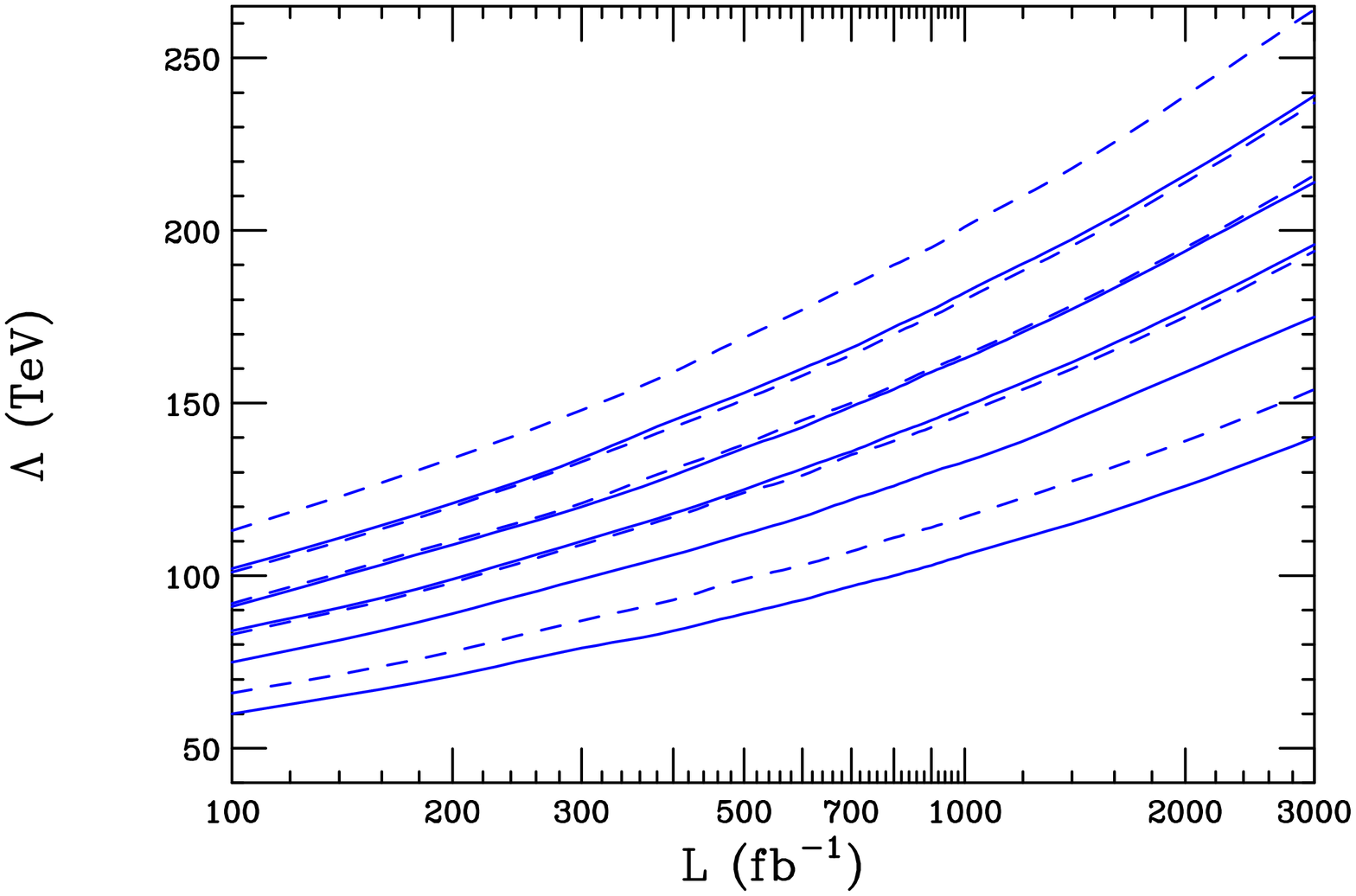}
\hspace*{5mm}
\includegraphics[width=5.6cm,angle=90]{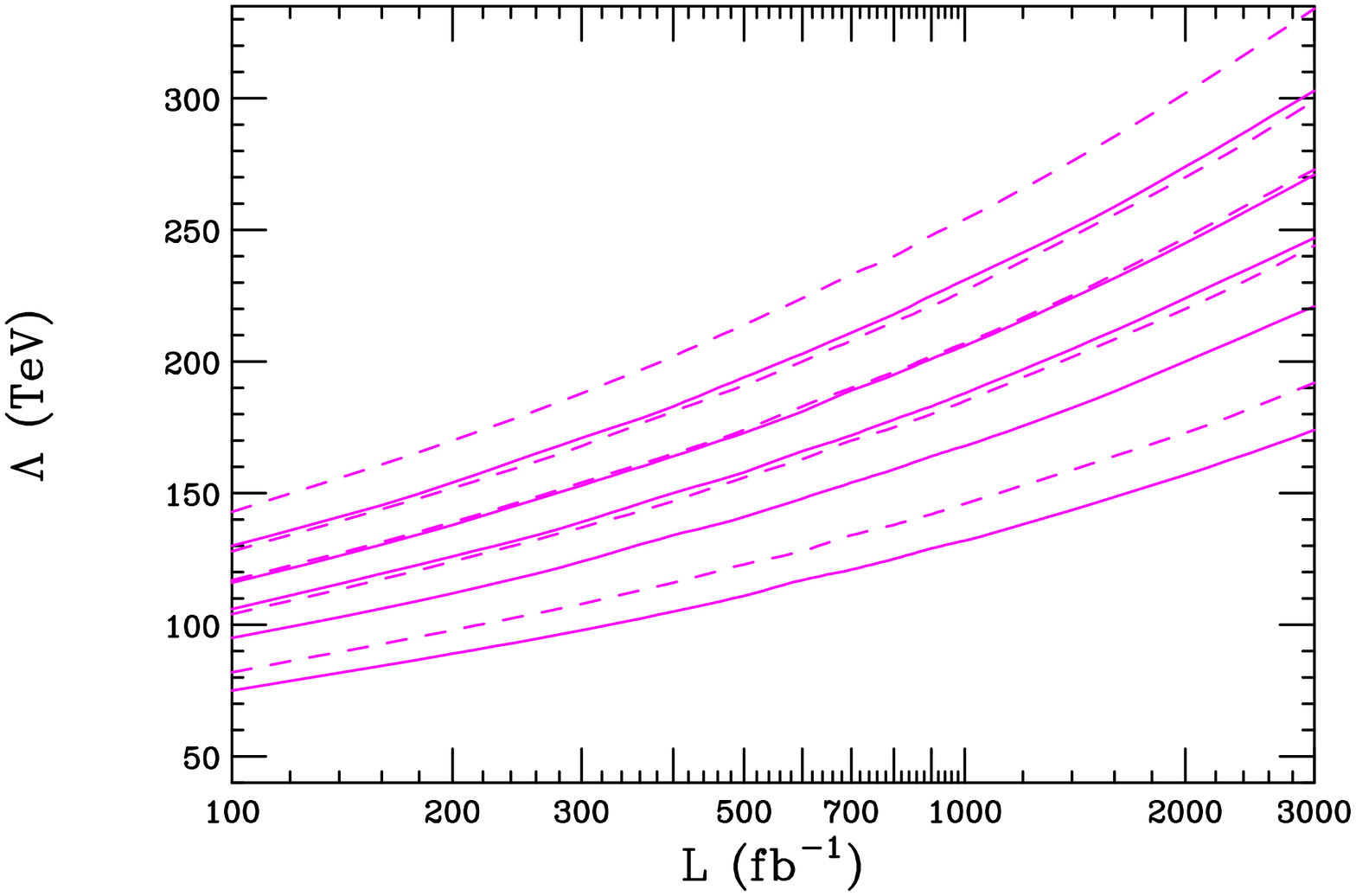}}
\vspace*{0.3cm}
\centerline{
\includegraphics[width=5.6cm,angle=90]{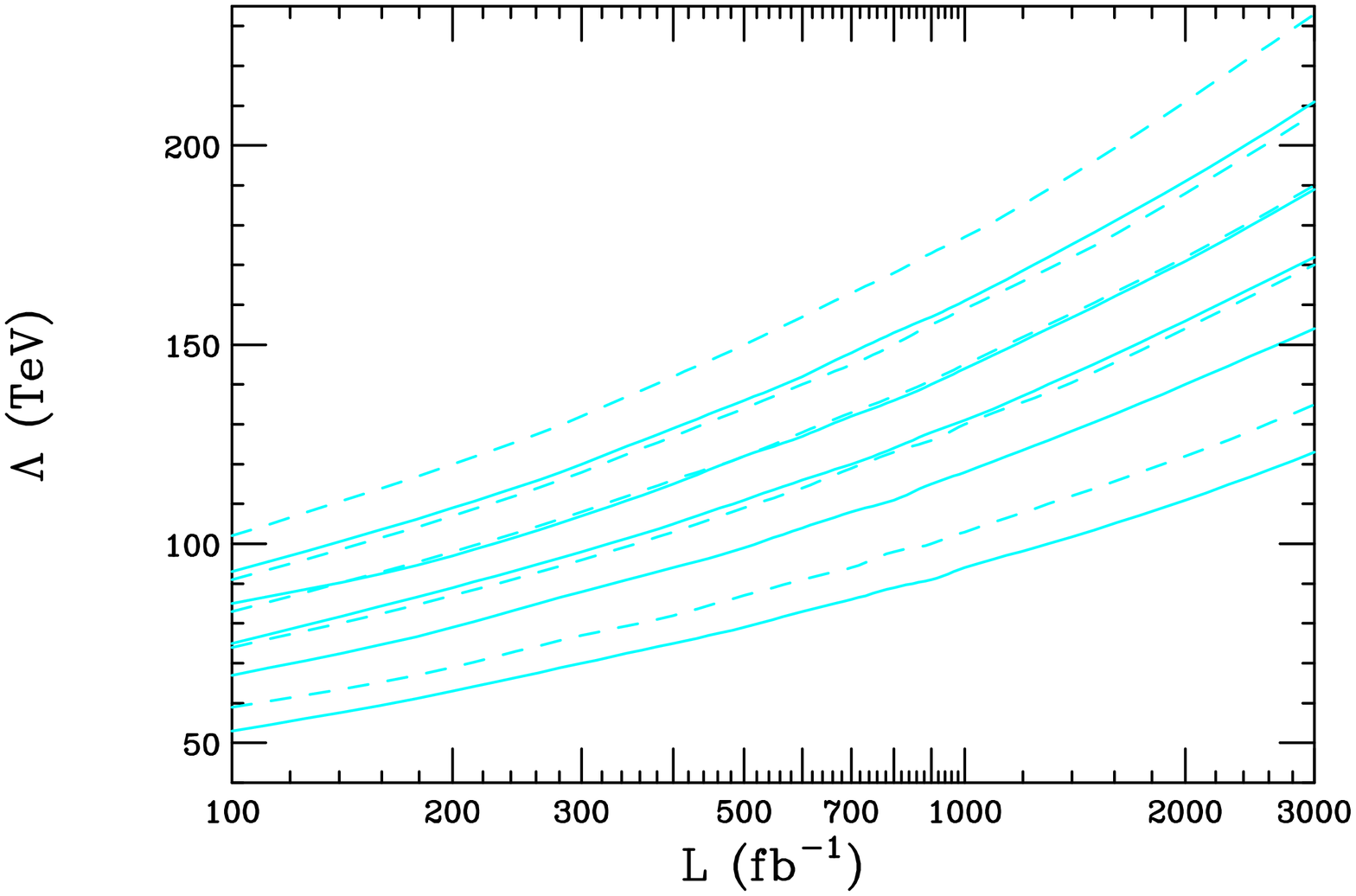}}
\vspace*{0.1cm}
\caption{Model dependence of the $95\%$ CL bounds on compositeness from Bhabha 
scattering. There are a pair of curves (solid,dashed) for each center 
of mass energy corresponding to $P_+=(0,60)\%$. From bottom to top in each 
panel the curves correspond to $\sqrt s=0.5,0.8,1,1.2$ and $1.5$ TeV. The top 
left(right) panel is for the LL(RR) case, the left(right) central panel is for 
the LR(VV) case and the lower panel is for the AA case, respectively.}
\label{fig8}
\end{figure}

\section{Tradeoffs and Prospects}

Given all of the results above for the different models 
we can now assess the various tradeoffs between variations 
in $\sqrt s$, $L$ and $P_+$ for contact interaction searches. Clearly 
graviton exchange in the ADD case 
will be distinct from the three other scenarios 
since it involves dimension-8 and not 
dimension-6 operators. To address these issues in the broadest possible way 
the curves shown in Figs.~\ref{fig9} and ~\ref{fig10} need to be used 
simultaneously. (Note that Fig.~\ref{fig10} shows that the growth in the search 
reach with increasing $P_+$ is roughly linear.) The 
results summarized in these figures rely on our conclusions from detailed 
calculations that the search reaches 
for contact interactions are at least approximately statistics dominated over the 
range of parameters of interest to us here. These are the most important results 
presented in this paper. 

To demonstrate the usage of these two sets of 
figures it is best to give a few examples. Suppose for the 
case of the LRM $Z'$ we want to know ($i$) the fractional 
increase in reach that is  obtained in going from $P_+=0$ to 
$P_+=50\%$ and what this additional reach would correspond to in 
terms of ($ii$) increased $L$ or ($iii$) increased $\sqrt s$? First, 
the left panel of  Fig.~\ref{fig10} 
tells us that in this case going to 
$P_+=50\%$ would lead to a search reach increase of about $15\%$ for the LRM. 
Now using the two top panels of 
Fig.~\ref{fig9} we see that a $15\%$ search reach gain from $P_+$ is 
equivalent to a $\sim 75\%$ increase in integrated 
luminosity or, instead,  a $\sim 32\%$ increase in 
$\sqrt s$ for any $Z'$ model. It is clear in this example that a significant 
amount of positron polarization buys you a lot in terms of equivalent 
luminosity or $\sqrt s$ increases. If we followed the same numerical 
example for the ADD model and repeated the last proceedure we 
would find an increased reach of only $5\%$ in going to $P_+=50\%$ 
which is equivalent to a $\sim 47\%$ 
increase in $L$ or a $\sim 7\%$ increase in $\sqrt s$. Here the gain 
from positron polarization is clearly somewhat less. 
We can also ask other questions (for fixed but arbitrary values of  
$P_+$), \eg , a factor of 5 increase in $L$ produces a search reach increase 
for the ADD model which is equivalent to how much of a corresponding increase 
in $\sqrt s$? From Fig.~\ref{fig9} we see this value is $\sim 31\%$. It is 
quite clear from these examples that a large number of issues regarding 
parameter options can now be addressed at least within the context of contact 
interaction searches. 

\begin{figure}[htbp]
\centerline{
\includegraphics[width=7.6cm,angle=90]{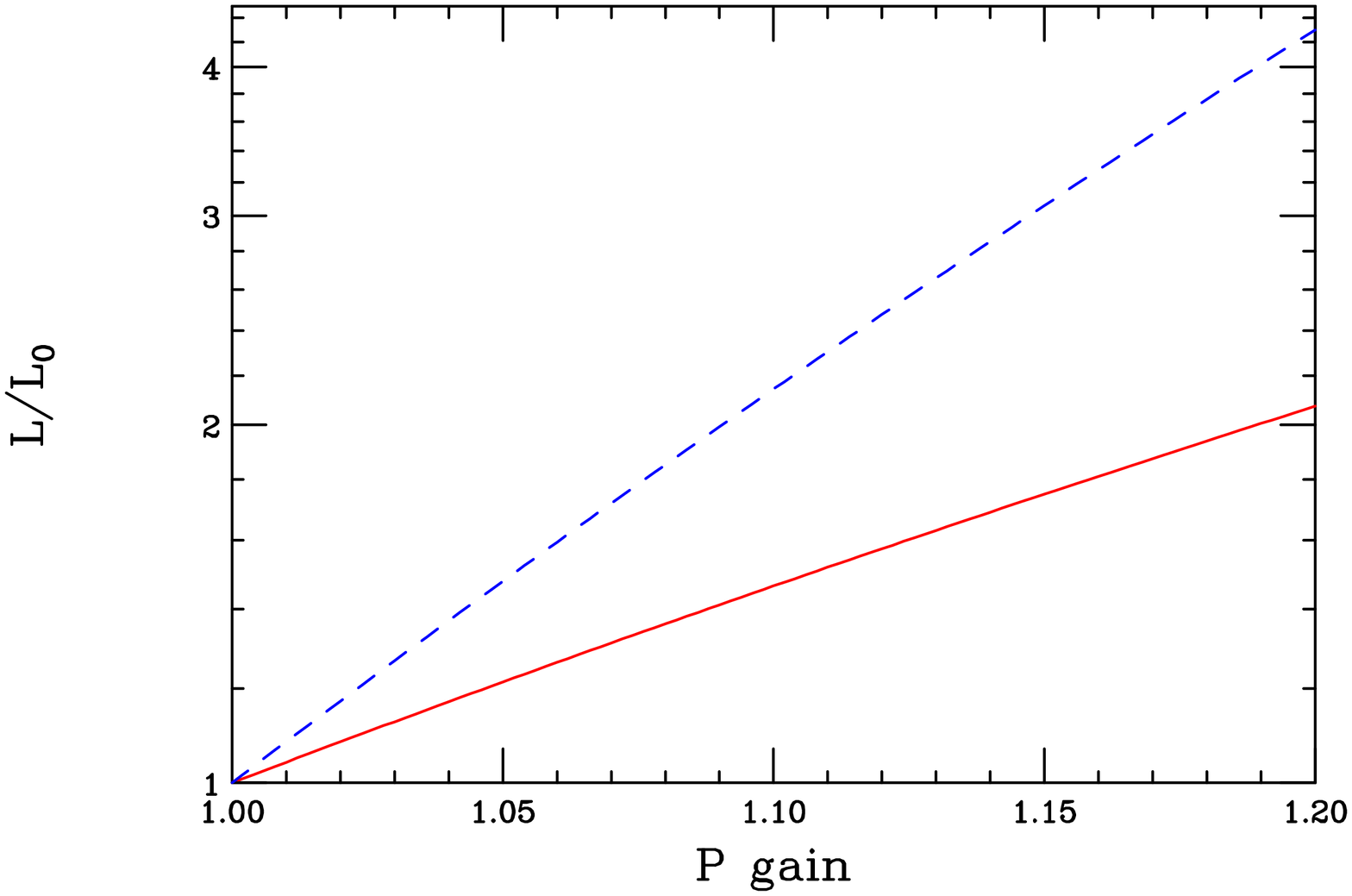}}
\vspace*{0.3cm}
\centerline{
\includegraphics[width=7.6cm,angle=90]{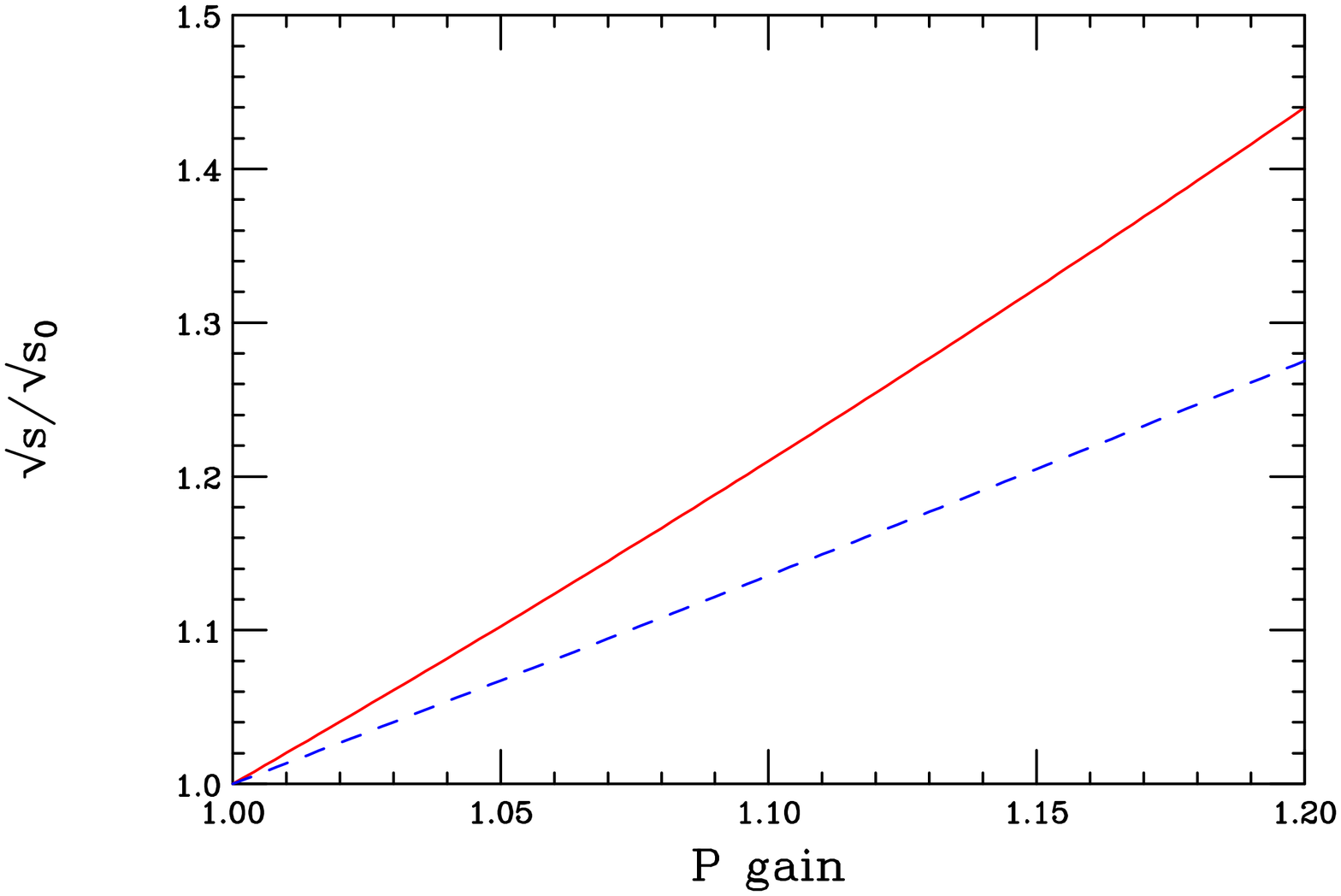}}
\vspace*{0.3cm}
\centerline{
\includegraphics[width=7.6cm,angle=90]{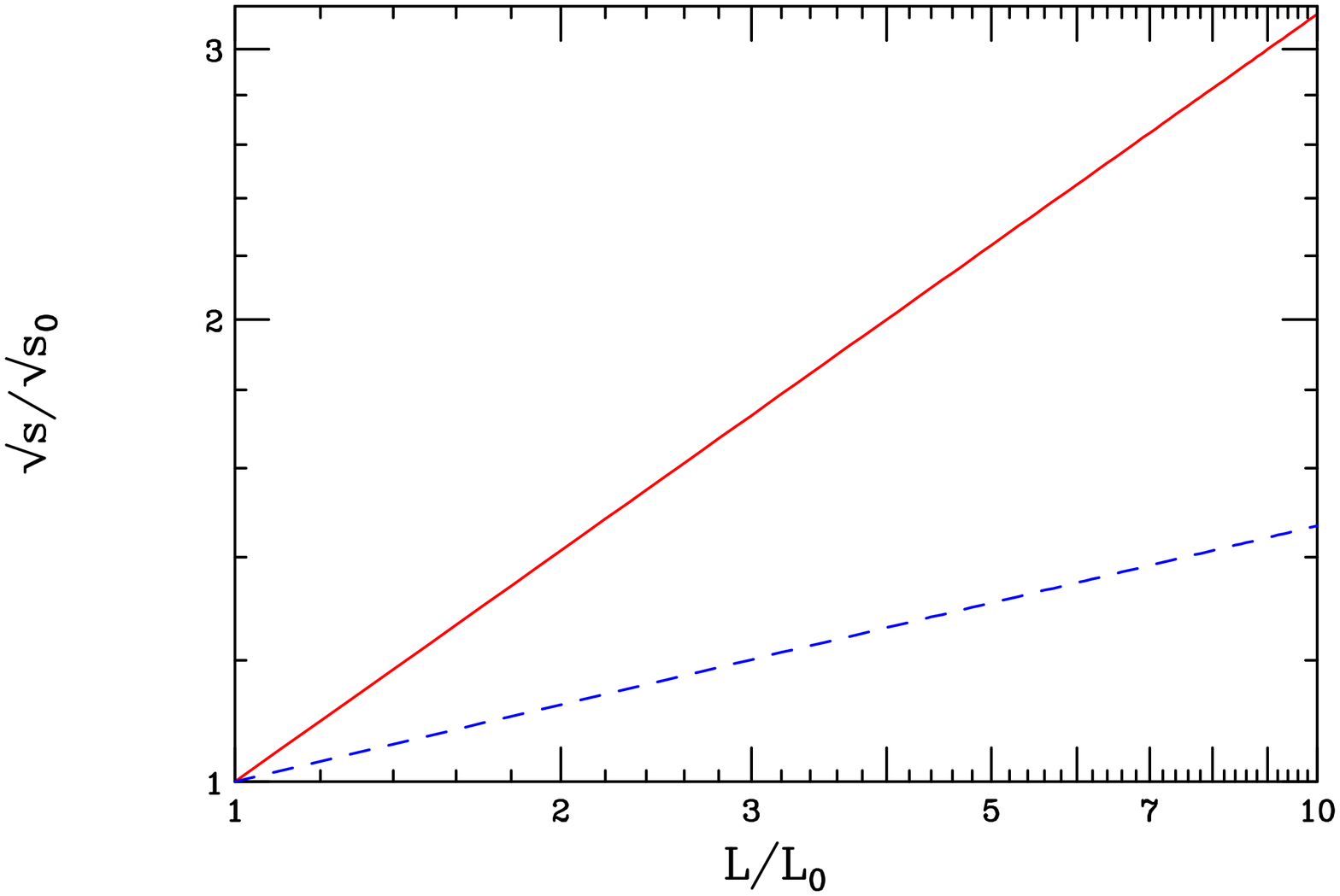}}
\vspace*{0.1cm}
\caption{Parameter tradeoffs for the $95\%$ CL search reaches for 
dimension-6(solid) and dimension-8(dashed) type contact interactions. 
`P gain' is the increase in search reach in going from $P_+=0$ to $P_+=60\%$ 
which can be read off the previous figures and is discussed in the text and shown 
in detail in the following Figure.}
\label{fig9}
\end{figure}

In planning for the LC it is important to explore what the 
physics benefits of different potential upgrade 
paths will be. At least in the case of contact interactions the detailed 
analysis above will hopefully be helpful in making the appropriate choices.

\begin{figure}[htbp]
\centerline{
\includegraphics[width=7.6cm,angle=90]{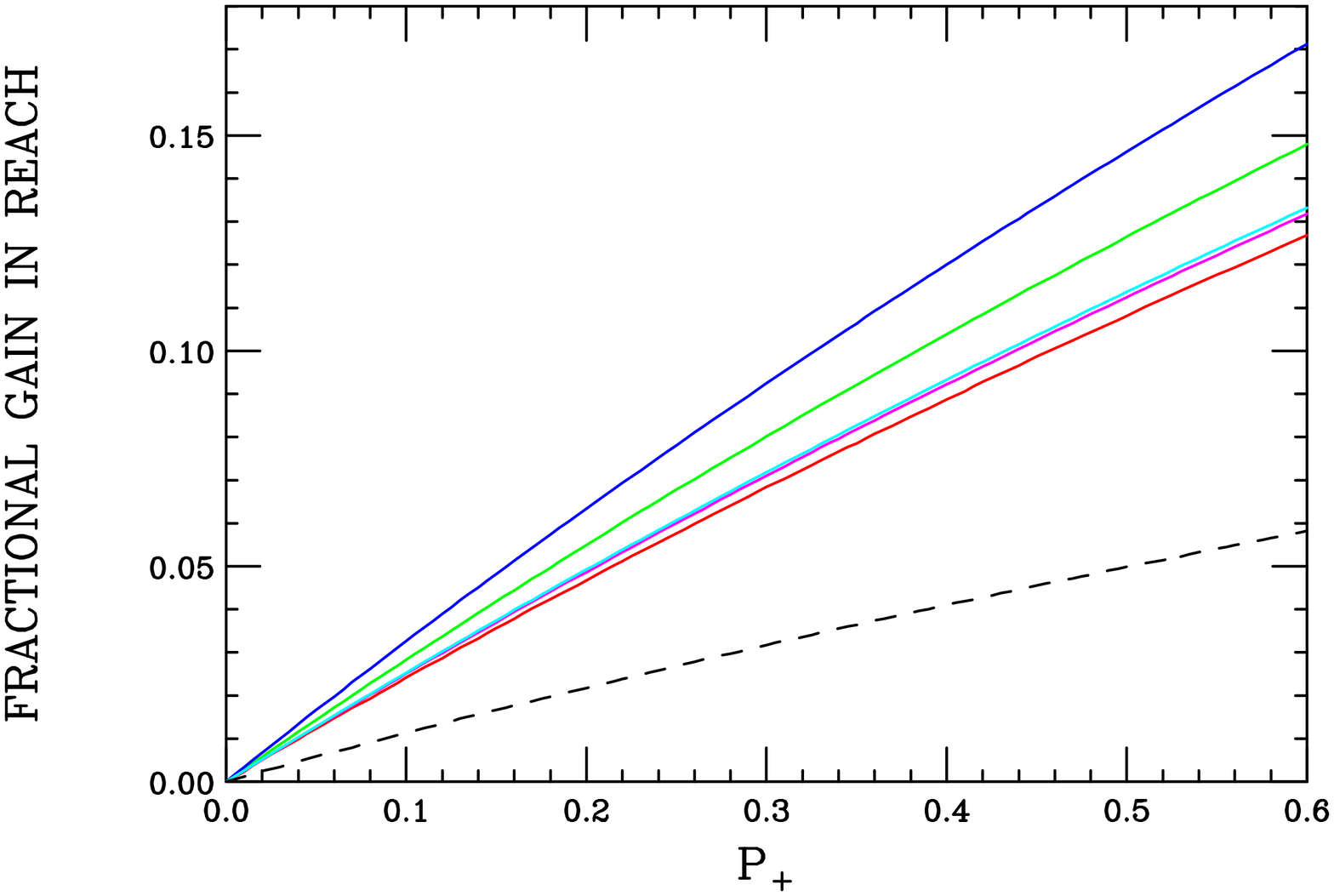}}
\vspace*{0.3cm}
\centerline{
\includegraphics[width=7.6cm,angle=90]{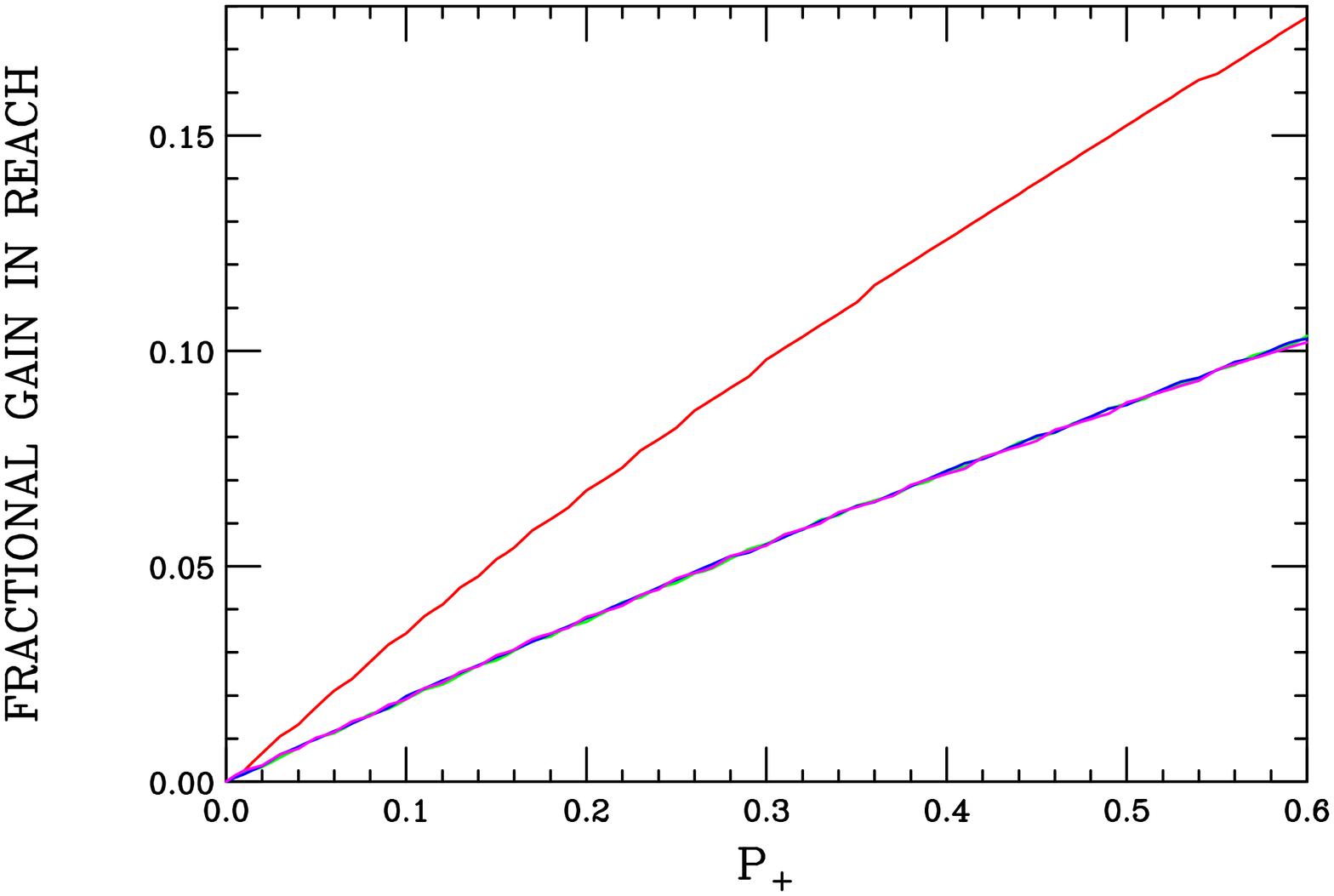}}
\vspace*{0.1cm}
\caption{Increase in search reach for fixed luminosity and $\sqrt s$ as a 
function of the amount of positron polarization. In the top panel are the 
results for the ADD model(dashed) as well as for $Z'$: $\psi$(green), 
$\chi$(red) and LRM(blue). The 5DSM and SSM correspond to the single 
violet curve. In the bottom panel are the corresponding results for 
compositeness 
searches in Bhabha scattering: the red curve is the result for the LL and RR 
cases while the second curve is for the LR,VV and AA cases.}
\label{fig10}
\end{figure}

\section{References}

\def\MPL #1 #2 #3 {Mod. Phys. Lett. {\bf#1},\ #2 (#3)}
\def\NPB #1 #2 #3 {Nucl. Phys. {\bf#1},\ #2 (#3)}
\def\PLB #1 #2 #3 {Phys. Lett. {\bf#1},\ #2 (#3)}
\def\PR #1 #2 #3 {Phys. Rep. {\bf#1},\ #2 (#3)}
\def\PRD #1 #2 #3 {Phys. Rev. {\bf#1},\ #2 (#3)}
\def\PRL #1 #2 #3 {Phys. Rev. Lett. {\bf#1},\ #2 (#3)}
\def\RMP #1 #2 #3 {Rev. Mod. Phys. {\bf#1},\ #2 (#3)}
\def\NIM #1 #2 #3 {Nuc. Inst. Meth. {\bf#1},\ #2 (#3)}
\def\ZPC #1 #2 #3 {Z. Phys. {\bf#1},\ #2 (#3)}
\def\EJPC #1 #2 #3 {E. Phys. J. {\bf#1},\ #2 (#3)}
\def\IJMP #1 #2 #3 {Int. J. Mod. Phys. {\bf#1},\ #2 (#3)}
\def\JHEP #1 #2 #3 {J. High En. Phys. {\bf#1},\ #2 (#3)}

\end{document}